\documentclass{iopart}
\usepackage{iopams,graphicx}
\begin{document}
\title{A new dissipation term for
finite-difference simulations in Relativity}
\author{Daniela Alic, Carles Bona and Carles Bona-Casas}
\address{Departament de Fisica, Universitat de les Illes Balears \\
Institute for Applied Computation with Community
Code (IAC$^3$).}
\begin{abstract}
We present a new numerical dissipation algorithm, which can be
efficiently used in combination with centered finite-difference
methods. We start from a formulation of centered finite-volume
methods for Numerical Relativity, in which third-order space
accuracy can be obtained by employing just piecewise-linear
reconstruction. We obtain a simplified version of the algorithm,
which can be viewed as a centered finite-difference method plus
some 'adaptive dissipation'. The performance of this algorithm is
confirmed by numerical results obtained from 3D black hole
simulations.
\end{abstract}
\section{Introduction}
In a recent paper (Alic \etal 2007), we presented a centered
finite-volume (CFV) method for black-hole simulations in numerical
relativity. This method is a variant of the well known
local-Lax-Friedrichs approach (LLF), which is currently being used
in computational fluid dynamics (including magneto-hydrodynamics).
For a specific choice of the parameters, this method can be
written as a piecewise-fourth-order finite-difference (FD)
algorithm plus a piecewise-third-order accurate artificial
dissipation, with automatically tuned local coefficient. The
piecewise prefix comes from the slope limiters that are
incorporated in order to deal with shocks or other
discontinuities.

Current black hole simulations in Numerical Relativity use instead
centered FD algorithms combined with a numerical dissipation term
of the Kreiss-Oliger type (Gustafson \etal 1995). This combination
can be interpreted as a single numerical scheme with built-in
dissipation, which can be tuned by a single parameter. In most
numerical relativity simulations, where only smooth profiles are
dealt with, this has shown to be an efficient computational
approach. In some black hole simulations, however, the required
amount of dissipation varies from the inner to the outer regions,
so this approach is lacking some flexibility.

Our main point is that, as far as the slope limiters are not
required, the FV algorithm which we developed can be expressed
also as a fourth-order centered FD algorithms combined with a
local dissipation term which is automatically adapted to the
requirements of the either interior or exterior black hole
regions.

\section{The Centered Finite-Volume Method in a Flux-Splitting Approach}

We consider the Einstein field equations written as a system of balance laws
\begin{equation}
\partial_{t}~ u + \partial_{k}~ F^{k}(u) = S(u)\,,
\end{equation}
where the Flux terms F and the Source terms S depend algebraically
on the array of dynamical fields u. We will consider first the
one-dimensional case. In a regular finite difference grid, we
choose the elementary cell to be the interval $(x_{i-1/2},
x_{i+1/2})$ centered in the grid point $x_{i}$. The resulting
discrete scheme is given by
\begin{equation}
u_{i}^{n+1} = u_{i}^{n} - \frac{\Delta t}{\Delta x}
~[~F_{i+1/2}^{x} - F_{i-1/2}^{x}~] + \Delta t \ S_{i}\,.
\end{equation}
This general algorithm requires the prescription of the interface
fluxes $F_{i \pm 1/2}^{x}$. We will use linear reconstruction: the
dynamical fields will be modelled as piecewise linear functions in
each cell.

Our CFD method (Alic \etal 2007) is based on the flux-splitting
approach, in which the information is evaluated at the grid nodes,
selecting the components of the flux that will propagate in each
direction. In every grid point, the flux can then be splitted into
two components
\begin{equation}
F_{i}^{\pm} =  F_{i} \pm \lambda_{i} u_{i}\,,
\end{equation}
where $\lambda$ is the maximum characteristic speed at the
specific grid point. In this way, we will have the freedom to
choose a different slope for each component, which will allow us
to improve the space accuracy.

Then we will consider at each interface two one-sided (left and
right) predictions from the neighboring points:
\begin{equation}
F_{L}^{+} = F_{i}^{+} + \frac{1}{2}~\sigma_{i}^{+}\,, \ \ \ \ \ \
F_{R}^{-} = F_{i+1}^{-} - \frac{1}{2}~\sigma_{i+1}^{-}\,,
\end{equation}
where $\sigma$ is the slope of the selected flux component in the
corresponding cell. The interface flux is obtained by recombining
both predictions
\begin{equation}
F_{i+1/2} =  \frac{1}{2}~(F_{L}^{+} + F_{R}^{-})\,.
\end{equation}

\section{Third-order-accurate Dissipation Formula}

Let us write the prescription for the slopes of the flux
components generically as
\begin{equation}
\sigma^{+}_{i} = a \ \sigma^{L}_{i} + (1-a) \ \sigma^{R}_{i}, \ \
\ \ \ \ \sigma^{-}_{i} = b \ \sigma^{L}_{i} + (1-b) \
\sigma^{R}_{i}\,,
\end{equation}
where $a$ and $b$ are slope coefficients, and we have noted for
short
\begin{equation}
\sigma^{L}_{i} = F_{i}^{\pm} - F_{i-1}^{\pm}\,, \ \ \ \ \ \
\sigma^{R}_{i} = F_{i+1}^{\pm} - F_{i}^{\pm}\,.
\end{equation}

\begin{figure}
\centering
\scalebox{0.5}[0.5]{\includegraphics{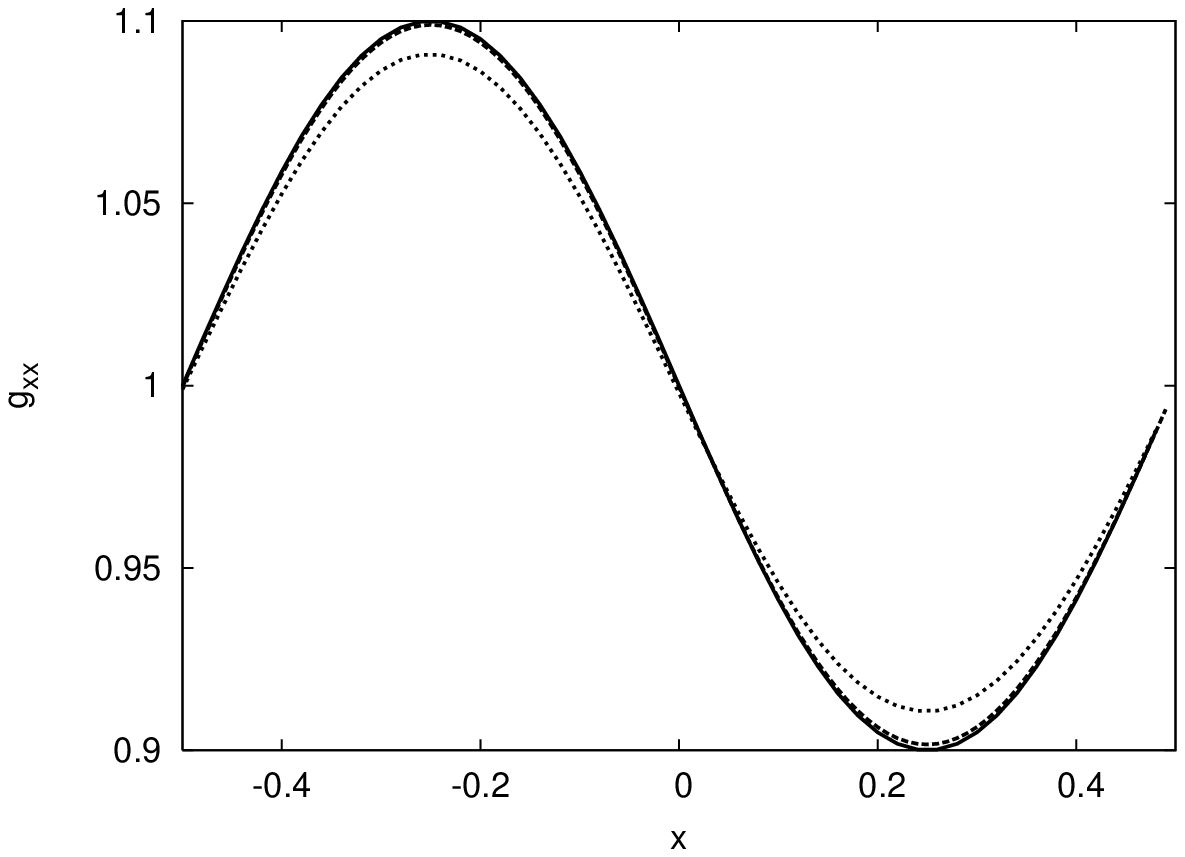}
\includegraphics{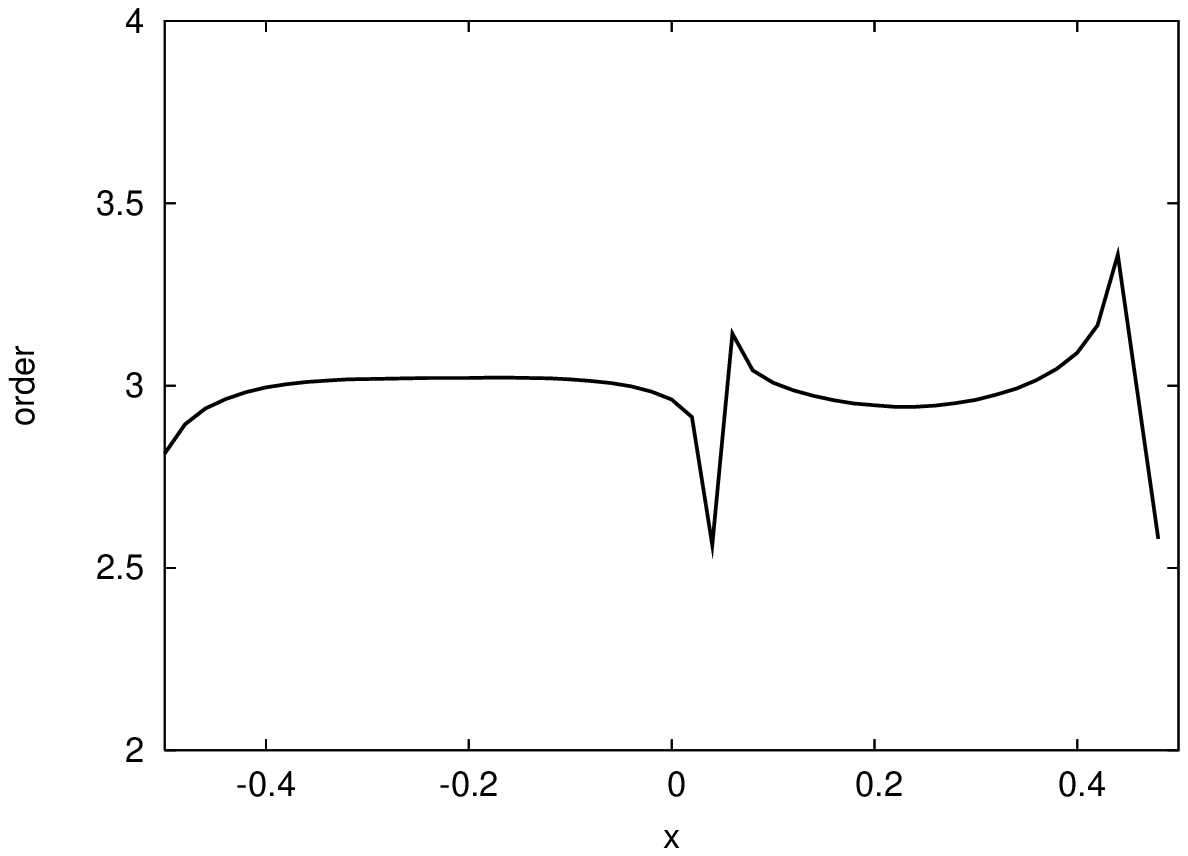}}
\caption{~3D Gauge-Wave test simulations. The profiles in the left
panel are plots of the metric component $g_{xx}$ for three
resolutions ($\Delta x = 0.02,\, 0.01,\, 0.005$) after 100
crossing times; the last two almost coincide. The right panel
shows the local convergence rate, calculated by comparing the two
higher resolutions with the exact solution, confirming third-order
convergence.}
\end{figure}

We determine the specific values of the slope coefficients which
allow third order accuracy, by inserting the slope formulae in the
CFV method. Comparing the final algorithm with the standard fourth
order finite difference method, one obtains the values $a=1/3$,
$b=2/3$ for the slope coefficients. If no slope limiters are
implemented, the derivative of the flux can be expressed in closed
form as
\begin{equation}
D_{x}(F_{i}) = \frac{1}{12\Delta x}~[- F_{i+2} + 8 \ F_{i+1} - 8 \
F_{i-1} + F_{i-2}~] + Dis(F_{i})\,,
\end{equation}
where the first part of the formula is just the centered
fourth-order FD algorithm and the second part is the new
dissipation term:
\begin{equation}\label{LLF}
\fl   Dis(F_{i}) = \frac{1}{12\Delta x}~[~\lambda_{i+2} u_{i+2} -
4\, \lambda_{i+1} u_{i+1} + 6\, \lambda_{i} u_{i} - 4\,
\lambda_{i-1} u_{i-1} + \lambda_{i-2} u_{i-2}~]\,.
\end{equation}

We test the stability and convergence of the resulting algorithm
in the Gauge Wave Test (Fig.~1), one of the standard tests for
Numerical Relativity (Alcubierre \etal 2004). The plots show that
the resulting amount of dissipation is actually very small and
confirm third-order convergence. Note, however, that in this case
all the local $\lambda$ coefficients are equal to one, so the new
dissipation term coincides with the Kreiss-Oliger one for a
specific value of its global coefficient.

\section{The 3D Black Hole}

The dissipation algorithm presented above can be easily extended
to the 3D case:
\begin{eqnarray}
&Dis(F^x_{i,j,k}) = \frac{1}{12\Delta x}~[~\lambda^x_{i+2,j,k} \
u_{i+2,j,k}- 4 \ \lambda^x_{i+1,j,k} \ u_{i+1,j,k} \nonumber\\
&+ 6 \ \lambda^x_{i,j,k} \ u_{i,j,k} - 4 \lambda^x_{i-1,j,k} \
u_{i-1,j,k} + \lambda^x_{i-2,j,k} \ u_{i-2,j,k}~]\,,
\end{eqnarray}
where $\lambda^x$ is the maximum characteristic speed along the
$x$ axis, and analogous formulae hold for the the $y$ and $z$
axes.

\begin{figure}[b]
\centering
\scalebox{1.5}[1.2]{\includegraphics[width=6cm,height=4cm]{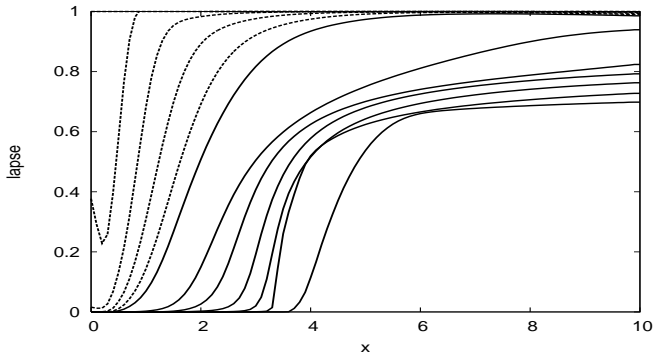}
} \caption{Lapse evolution in a 3D black hole simulation (zero
shift). The dotted line profiles are plotted every 1M. The solid
line ones are plotted every 5M, up to 35M, before boundary-related
features become too important (the boundary is just at 10M).}
\end{figure}

Let us consider initial data taken from a Schwarzschild black hole
\begin{equation}\label{eq::Schwarzschild}
ds^{2} = - \alpha^{2}dt^{2} + (1
+\frac{M}{2r})^{4}~\delta_{ij}~dx^idx^j\,.
\end{equation}
(isotropic coordinates). We will use the 'stuffed black hole'
approach (Arbona etal 1998), by matching a scalar field interior
metric to (\ref{eq::Schwarzschild}) (the scalar field will also
evolve). As gauge conditions we choose a singularity-avoidant
slicing of the '1+log' type in normal coordinates (zero shift).

We present in (Fig. 2) a low-resolution simulation ($\Delta x =
0.1M$) which proves the performance of our numerical method in 3D
strong-field scenarios. Even in presence of steep gradients, the
lapse profiles evolve smoothly.

The numerical tests shown here have been performed with the Z3
evolution system (first order in space and time) written in a flux
conservative form (Bona \etal 1989). The time integration is dealt
by the well-known method-of-lines, with a third-order Runge-Kutta
algorithm.

\section*{Acknowledgements}
This work has been supported by the Spanish Ministry of Science
and Education, through FPI and FPU fellowships and the research
grant FPA2004-03666, and by the Balearic Conselleria d'Economia
Hissenda i Innovaci\'{o} through the grant PRDIB-2005GC2-06.

\References

\item[] Alcubierre, M. \etal\  2004, {\it Class. Quantum Grav.},
21(2), 589613.

\item[] Alic, D., Bona, C., Bona-Casas,
   C., Masso, J., 2007, {\it Phys. Rev. D} 76, 104007.

\item[] Arbona, A. \etal\ 1998, {\it Phys. Rev. D} 57, 2397.

\item[] Bona, C., Masso, J., 1989, {\it Phys. Rev. D} 40, 1022.

\item[] Gustafson, B., Kreiss, H.O., Oliger, J.,
   1995, {\it Time dependent problems and difference methods}, Wiley, New York.

\endrefs
\end{document}